\documentstyle[12pt,cite]{article}

\newcommand{\be}{\begin{equation}}
\newcommand{\ee}{\end{equation}}
\def\bea{\begin{eqnarray}}
\def\eea{\end{eqnarray}}
\def\nn{\nonumber \\}

\def\x{x_{\perp}}
\def\y{y_{\perp}}
\def\k{k_{\perp}}
\def\lapprox{\lower .7ex\hbox{$\;\stackrel{\textstyle <}{\sim}\;$}}
\def\gapprox{\lower .7ex\hbox{$\;\stackrel{\textstyle >}{\sim}\;$}}

\input psfig

\begin{document}
\date{}
\title{
{\large\rm DESY 98-113}\hfill{\large\tt ISSN 0418-9833}\\
{\large\rm DAMTP-1998-113}\hfill\vspace*{0cm}\\
{\large\rm August 1998}
\hfill\vspace*{2.5cm}\\
Inclusive and Diffractive\\
Structure Functions at Small $x$}
\author{W. Buchm\"uller, T. Gehrmann\\
{\normalsize\it Deutsches Elektronen-Synchrotron DESY, 22603 Hamburg, 
Germany}
\\[.2cm]
and\\[.2cm]
A. Hebecker\\
{\normalsize\it D.A.M.T.P., Cambridge University, Cambridge CB3 9EW, 
England}
\vspace*{2cm}\\                     
}

\maketitle  
\begin{abstract}
\noindent
In the semiclassical approach, inclusive and diffractive quark and gluon 
distributions are expressed in terms of correlation functions of Wilson 
loops. Each Wilson loop integrates the colour field strength in the area 
between the trajectories of two fast partons penetrating the proton. We
introduce a specific model for averaging over the relevant colour field 
configurations. Within this model, all parton distributions at some low 
scale $Q_0^2$ are given in terms of three parameters. Inclusive and 
diffractive structure functions at higher values of $Q^2$ are determined 
in a leading-order QCD analysis. In both cases, the evolution is driven by 
a large gluon distribution. A satisfactory description of the structure 
functions $F_2(x,Q^2)$ and $F_2^{D(3)}(\xi,\beta,Q^2)$ is obtained. The 
observed rise of $F_2^{D(3)}$ with $\xi$ is parametrized by a 
non-perturbative logarithmic energy dependence, compatible with unitarity. 
In our analysis, the observed rise of $F_2$ at small $x$ is largely due to 
the same effect. 
\end{abstract} 
\thispagestyle{empty}
\newpage

\section{Introduction}

Recently, precise measurements of inclusive \cite{incl} and diffractive 
\cite{diff} structure functions at small $x$ have become available. Numerical 
analyses, based on different theoretical approaches, have been performed by 
many authors (see e.g.~\cite{nz,num,gs,bekw}). 

In the present paper, a joint analysis of diffractive and inclusive 
structure functions based on the semiclassical calculations of \cite{bhm,h} 
is performed. The idea of a close similarity between diffractive and 
non-diffractive processes in deep inelastic scattering (DIS) lies at the 
heart of the semiclassical approach. In both cases, a partonic fluctuation 
of the incoming virtual photon scatters off a superposition of target 
colour fields. If the scattered partons emerge in an overall colour singlet 
configuration, a diffractive final state results. 

Comparing the semiclassical description of structure functions with 
parton model expressions, we define 
inclusive and diffractive~\cite{dpd} parton distributions in 
the semiclassical approach. Higher order contributions in the semiclassical 
calculation exactly reproduce
the leading logarithmic corrections to the parton model, showing the 
consistency of both approaches. 
The semiclassical method is therefore used 
at some low scale to derive initial distributions. Starting from these 
distributions, a leading-order DGLAP analysis~\cite{dglap} of experimental 
data is performed. 

The calculation of the above initial distributions involves averaging over 
all relevant colour field configurations. To perform this averaging, a 
simple non-perturbative model, valid in the case of large hadronic targets 
\cite{mv}, is used. It is based on the observation that, for 
extended target colour fields, the transverse size of partonic fluctuations 
of the photon remains small \cite{hw}. 

The energy dependence arising from the large-momentum cutoff applied in 
the process of colour field averaging can not be calculated from first 
principles. It is described by a $\ln^2 x$ ansatz, consistent with unitarity. 
In the semiclassical approach, this energy dependence is expected to 
be universal for both the inclusive and diffractive 
structure functions~\cite{b}. 

Overall, a satisfactory description of inclusive and diffractive small-$x$ 
structure functions, based on a minimal number of assumptions and only four 
fitted parameters, is achieved. In our opinion, this lends support to the 
idea of a close similarity between the mechanisms of inclusive and 
diffractive scattering and its implementation in the semiclassical approach. 

The paper is organized as follows. In Sections \ref{f2sec} and \ref{f2dsec}, 
semiclassical formulae for inclusive and diffractive parton distributions 
are given, and the underlying physical picture is discussed. Section~\ref{av} 
deals with our model for the colour field averaging that is 
responsible for the input distributions. The leading-order DGLAP analysis 
and the comparison with experimental data are the subject of Sect.~\ref{na}, 
followed by conclusions in Sect.~\ref{conc}. Appendix A contains some 
additional formulae relevant for the calculation of the diffractive parton 
distribution functions. Finally, we illustrate in Appendix B how the 
diffractive parton distributions of a small colour dipole, calculated in 
\cite{hks}, can be obtained using the semiclassical method.

\section{Inclusive structure functions}\label{f2sec}

\begin{figure}[t]
\begin{center}
\vspace*{-.5cm}
\parbox[b]{13.7cm}{\psfig{width=13.7cm,file=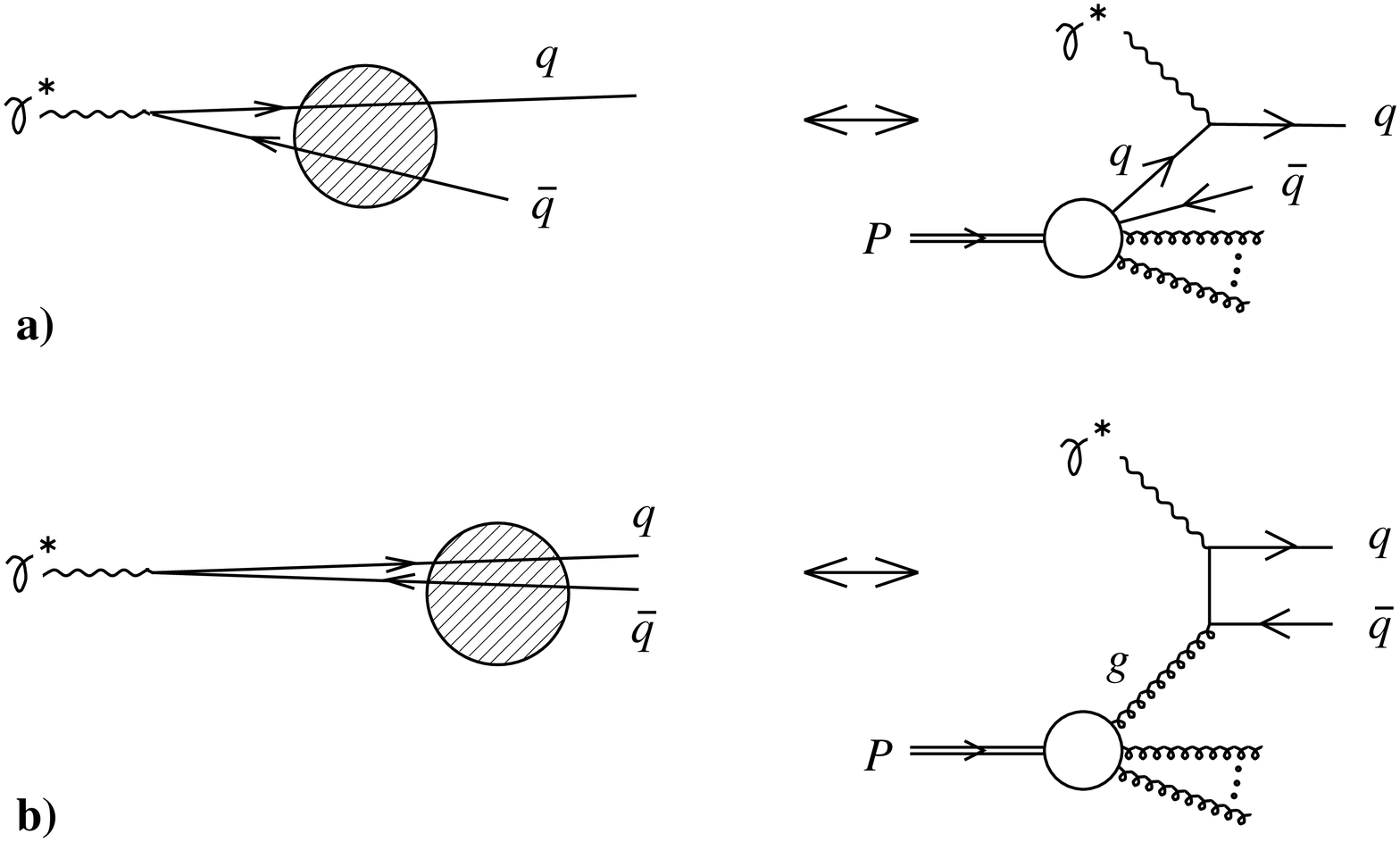}}\\
\end{center}
\refstepcounter{figure}
\label{f2}
{\bf Figure \ref{f2}:} Inclusive DIS in the proton rest frame (left) and the 
Breit frame (right); asymmetric fluctuations correspond to quark scattering 
(a), symmetric fluctuations to boson-gluon fusion (b). 
\end{figure}

Small-$x$ DIS can be conveniently discussed in terms of the $q\bar{q}$ wave 
function of the virtual photon (see e.g. \cite{nz}). In the semiclassical 
approach, the corresponding $q\bar{q}$ states scatter off a `soft' target 
colour field (cf. the l.h. side of Fig.~\ref{f2}). As $x\rightarrow 0$, the 
resulting inclusive structure function $F_T(x,Q^2)$ approaches a constant 
\cite{bhm}, 
\bea
F_T(x,Q^2)&=& {2Q^2 e_q^2\over (2\pi)^4}\int_0^1 d\alpha (\alpha^2
             +(1-\alpha)^2) N^2 \int_{\y} K_1(y N)^2 
              \int_{\x} \mbox{tr}\left(W^{\cal F}_{\x}(\y)
              W_{\x}^{{\cal F} \dagger}(\y)\right)
\nonumber\\
\nonumber\\
&&+\,{\cal O}(x)\,.\label{ftsc}
\eea
The light-like paths of quark and antiquark penetrate the colour field of 
the proton at transverse positions $\x$ and $\x + \y$ picking up non-Abelian 
phase factors $U^{\cal F}(\x)$ and $U^{{\cal F}\dagger}(\x+\y)$ (where 
${\cal F}$ stands for the fundamental representation). The function 
\be
W^{\cal F}_{\x}(\y)=U^{\cal F}(\x)U^{{\cal F}\dagger}(\x+\y)-1\label{wuu}
\ee
is essentially a closed Wilson loop through the corresponding section of 
the proton, which measures an integral of the proton colour field strength. 
Furthermore, $\alpha$ is the fraction of the photon momentum carried by 
the quark, $N^2 = \alpha (1 - \alpha) Q^2$, $y=|\y|$ and $e_q$ is the quark 
charge. 

To map this calculation onto the conventional parton model framework, 
extract the leading-twist contribution from Eq.~(\ref{ftsc}) and identify it 
with $F_T(x,Q^2)=2e_q^2xq(x,Q^2)$. The resulting quark distribution reads 
\bea
xq(x,Q^2) &=&{2\over (2\pi)^4}\int_0^{\mu^2} N^2\,dN^2 \int_{\y} K_1(y N)^2 
              \int_{\x} \mbox{tr}\left(W^{\cal F}_{\x}(\y)
              W_{\x}^{{\cal F} \dagger}(\y)\right)
\nonumber\\
\nonumber\\
            && +{2\over 3(2\pi)^3}\left(\ln\frac{Q^2}{\mu^2}-1\right)
              \int_{\x} \mbox{tr}\left(\partial_{\y}W^{\cal F}_{\x}(0)
              \partial_{\y}W_{\x}^{{\cal F} \dagger}(0)\right)\;.\label{qi}
\eea
Note that this expression is independent of $\mu^2$, which has only been 
introduced to separate the logarithmic $Q^2$ dependence explicitly. We have 
assumed that the hadronic structure, encoded in the $y^2$ dependence of 
$WW^\dagger$, is dominated by the soft scale $\Lambda^2$, which satisfies 
$\Lambda^2\ll\mu^2\ll Q^2$.

The corresponding gluon distribution at small $x$ is most easily calculated as
\be
xg(x,Q^2) = \frac{3\pi}{\alpha_s e_q^2}\,\cdot\,\frac{\partial F_T(x,Q^2)}
{\partial\ln Q^2} = {1\over 2\pi^2 \alpha_s}\int_{\x} \mbox{tr}\left(
\partial_{\y}W^{\cal F}_{\x}(0)\partial_{\y}W_{\x}^{{\cal F} \dagger}(0)\right)
\,.\label{gi}
\ee
Equations (\ref{qi}) and (\ref{gi}) provide the basis for our analysis of 
inclusive DIS. Note that the semiclassical approach predicts  $g(x,Q^2) 
\sim 1/x$, which is expected for a classical bremsstrahlung spectrum of 
gluons.

To gain more physical insight into the correspondence of the semiclassical 
and the parton model approach, return to our starting point, 
Eq.~(\ref{ftsc}). It is instructive to view $F_T$ as a sum of two terms: 
$F_T^{asym}$, the contribution of asymmetric configurations where quark or 
antiquark are slow, $\alpha < \mu^2/Q^2$ or $1-\alpha < \mu^2/Q^2$ 
(Fig.~\ref{f2}a), and $F_T^{sym}$, the contribution of symmetric 
configurations where both quark and antiquark are fast, $\alpha, 1-\alpha > 
\mu^2/Q^2$ (Fig.~\ref{f2}b). In an infinite momentum frame of the proton, 
the asymmetric and symmetric contribution to $F_T$ correspond to 
photon-quark scattering and photon-gluon fusion respectively. 

The symmetric part is dominated by small $q\bar{q}$ pairs, i.e., by the short 
distance contribution to the Wilson-loop trace, 
\be 
\int_{\x} \mbox{tr}\left(W^{\cal F}_{\x}(\y)
              W_{\x}^{{\cal F} \dagger}(\y)\right)
= {1\over 2} y^2 \int_{\x} \mbox{tr}\left(\partial_{\y}W^{\cal F}_{\x}(0)
              \partial_{\y}W_{\x}^{{\cal F} \dagger}(0)\right)
  +{\cal O}(y^4)\; .
\ee
The corresponding contribution to the structure function is related to the 
second term on the r.h. side of Eq.~(\ref{qi}), which generates the gluon 
distribution, Eq.~(\ref{gi}). It was evaluated 
in \cite{bhm} at leading order in $\Lambda^2/\mu^2$ and $\mu^2/Q^2$,
\be
F_T^{sym}(0,Q^2) = {e_q^2\over 2\pi^3} \int_0^1 dz P_{qg}(z)
\left(\ln{Q^2\over \mu^2} - 1\right)
\int_{\x} \mbox{tr}\left(\partial_{\y}W^{\cal F}_{\x}(0)
\partial_{\y}W_{\x}^{{\cal F} \dagger}(0)\right)\,.
\ee
Here $P_{qg}(z)$ is the conventional gluon-quark splitting function. 

The other splitting functions appear if $\alpha_s$ corrections to $F_T$ and
$F_L$, associated with higher Fock states of the virtual photon, are 
considered in the semiclassical approach. For example, the $q\bar{q}g$ 
parton configuration involves, in the case where one of the quarks carries 
a small fraction of the photon momentum, a $\ln Q^2$ term associated with 
$P_{qq}(z)$.

The splitting function $P_{gg}(z)$ is most easily derived by considering 
an incoming virtual scalar which couples directly to the gluonic action 
term $F_{\mu\nu}F^{\mu\nu}$. Its lowest order Fock state consists of two 
gluons. We have checked explicitly that the semiclassical calculation of 
the corresponding high energy scattering process yields the usual 
gluon-gluon splitting function. 

Having shown that the leading logarithmic QCD corrections to the 
semiclassical approach exactly reproduce the well-known DGLAP splitting 
functions~\cite{dglap}, we do not pursue the calculation of higher order 
$\alpha_s$ contributions along the lines discussed above. Instead, the large 
logarithms $\ln\left(Q^2/\mu^2\right)$ are resummed in the conventional way, 
by means of the renormalization group. To this end, the parton distributions 
$q(x,Q^2)$ and $g(x,Q^2)$ are evaluated using DGLAP evolution equations, 
with the input distributions $q(x,Q_0^2)$ and $g(x,Q_0^2)$ given by 
Eqs.~(\ref{qi}) and (\ref{gi}). Here $Q_0^2$ is some small scale where 
logarithmic corrections are not yet important. 
The parton model description of the structure function at
leading order includes only photon-quark scattering. The leading 
logarithmic term from 
the photon-gluon fusion process appears now as part of the resummed 
quark distribution. 

So far, we have considered electroproduction off a fixed `soft' colour 
field. As a consequence, the unevolved structure function $F_T(x,Q_0^2)$ 
approaches a constant value as $x\rightarrow 0$. However, a proper treatment 
of the target requires the integration over all relevant colour field 
configurations. 

The qualitative features of the field averaging procedure are most 
conveniently discussed for the simple case of $q\bar{q}$ pair production 
in the proton rest frame. Consider the corresponding amplitude in a 
`mixed' representation, $<\,q\bar{q}\,A\,|\,\gamma^*p\,>$, where the final 
state consists of the outgoing $q\bar{q}$ pair and a colour field 
configuration $A$. We neglect any time evolution of the field between the 
actual scattering process and the moment at which the final state field 
configuration $A$ is defined. The squared amplitude, summed over all fields 
$A$ and normalized to the total space-time volume, reads 
\be
\frac{1}{VT}\,\int_A |<\,q\bar{q}\,A\,|\,\gamma^*p\,>|^2=4\pi m_p\delta(k_q^0
+k_{\bar{q}}^0-q^0)\int_{A_{loc}}\,\Big|\,\Phi_p[A_{loc}]\,F[A_{loc}]\,
\Big|^2\,.\label{ts}
\ee
Here $\Phi_p$ is the proton wave functional, $F$ is defined by the amplitude 
for the scattering off a fixed field $A$, 
\be
<\,q\bar{q}\,|\,\gamma^*\,>_A=2\pi\delta(k_q^0+k_{\bar{q}}^0-q^0)\,F[A]
\,,
\ee
$m_p$ is the proton mass, and $q$, $k_q$, $k_{\bar{q}}$ are the momenta of 
the incoming photon and the outgoing quark and antiquark respectively. The 
index `$loc$' symbolizes that, on the r.h. side of Eq.~(\ref{ts}), the 
integration is restricted to fields localized at, say, $\vec{x}=0$. This 
can be justified using translation covariance of the proton wave functional 
and of the matrix element $<\,q\bar{q}\,|\,\gamma^*\,>_A$ (cf. Sect.~2 of 
\cite{bhm}).

When writing $W^{\cal F}_{\x}(\y)$ we have, until now, always assumed the 
functional dependence on the classical colour field configuration $A_{cl}$ 
to be 
implicit, so that one should really read $W^{\cal F}_{\x}(\y)[A_{cl}]$. As 
can be seen from Eq.~(\ref{ts}), the full inclusive parton distributions are 
obtained from the previous formulae by the substitution 
\be
\mbox{tr}\!\left(W^{\cal F}_{\x}(\y)[A_{cl}]\,W^{{\cal F}\dagger}_{\x}(\y)
[A_{cl}]\right)\rightarrow\int_{A_{loc}}\!\Big|\,\Phi_p[A_{loc}]\,\Big|^2\,\,
\mbox{tr}\!\left(W^{\cal F}_{\x}(\y)[A_{loc}]\,W^{{\cal F}\dagger}_{\x}(\y)
[A_{loc}]\right).
\ee
The same applies to the diffractive distributions of the next section. An 
explicit model for the field average will be described in 
Sect.~\ref{av}. 

Decomposing the field $A_{loc}$ in Eq.~(\ref{ts}) into its Fourier modes 
$\tilde{A}_{loc}(\vec{k})$, the path integral can be written as
\be
\int_{A_{loc}} = \prod_{|\vec{k}|\ll|\vec{q}|}\int d\tilde{A}_{loc}(\vec{k})
\,,
\ee
where the cutoff $|\vec{q}|$ is required to ensure that the basic 
precondition 
for the semiclassical treatment, the softness of the target colour field 
with respect to the momenta of the fast particles, is respected. This cutoff 
induces a non-trivial energy dependence of the squared amplitude in 
Eq.~(\ref{ts}). 

We are not able to derive the explicit form of that energy dependence from 
first principles. Nevertheless, based on the above qualitative discussion, 
we ascribe a soft, non-perturbative energy growth to 
our input parton distributions used in the numerical analysis of 
Sect.~\ref{na}.

\section{Diffractive structure functions}\label{f2dsec}

The diffractive cross sections obtained in the
semiclassical approach \cite{bhm} can be expressed as convolution of the 
ordinary partonic cross sections and diffractive parton distributions, 
calculated in \cite{h}. The $q\bar{q}$ and $q\bar{q}g$ configurations then 
yield 
\bea
\frac{d\sigma_{T,L}}{d\xi} & = & \int_x^{\xi} dy\,\Bigg\{ \left( 
\hat{\sigma}_{T,L}(y)^{\gamma^*q\to q} 
+ \hat{\sigma}_{T,L}(y)^{\gamma^*
q\to qg} \right) \; \frac{dq(x/y,\xi)}{d\xi} \nonumber \\
& & \hspace{1.15cm} +\; \hat{\sigma}_{T,L}(y)^{\gamma^*g \to q\bar{q}}
\; \frac{dg(x/y,\xi)}{d\xi} \Bigg\}\;.
\eea
Here $\xi$ is related to the diffractive mass $M$ by $\xi=x_{I\!\!P}=x/ 
\beta$ and $\beta=Q^2/(Q^2+M^2)$. The corresponding structure functions 
read 
\bea
F_T^D(\xi,\beta,Q^2) &=& 2e_q^2x \int_{\beta}^1 {db\over b} \Bigg\{\left(
\delta(1-z) + {\alpha_s\over 2\pi}\left(P_{qq}(z)\ln{Q^2\over \mu^2} + 
\ldots\right) \right){dq(b,\xi)\over d\xi}\nonumber\\
&&\hspace{1cm} + {\alpha_s\over 2\pi}\left(P_{qg}(z)\ln{Q^2\over \mu^2}
+\ldots\right){dg(b,\xi)\over d\xi}\Bigg\}\;,\label{ftd}\\
F_L^D(\xi,\beta,Q^2) &=& 2e_q^2x {\alpha_s\over 2\pi} 
\int_{\beta}^1 {db\over b} \Bigg\{2C_F z {dq(b,\xi)\over d\xi}
+ 4T_F z(1-z) {dg(b,\xi)\over d\xi}\Bigg\}\;,\label{fld}
\eea
where $z=\beta/b$, and $C_F$ and $T_F$ are the usual colour factors. The 
physical interpretation of the diffractive parton distributions in the Breit 
frame is analogous to the interpretation of the inclusive distributions. 
The function $df(b,\xi)/d\xi$ (where $f=q,g$) is a conditional probability 
distribution. It describes the probability of finding a parton $f$, carrying 
a fraction $\xi b$ of the proton momentum, inside a colour-neutral cluster 
of partons that carries in total a fraction $\xi$ of the proton momentum. 

From Eqs.~(\ref{ftd}) and (\ref{fld}), it is obvious that the diffractive 
structure functions satisfy the ordinary DGLAP evolution equations (cf. the 
more general discussion of \cite{c}). The perturbative evolution takes place 
in the variables $\beta$ and $Q^2$; $\xi$ acts merely as a parameter. The 
physical reason for this is intuitively clear: for an arbitrary DIS event 
the invariant hadronic mass is $W$, and the quark which couples to the 
virtual photon can be radiated by a parton whose fraction of the proton 
momentum varies from 1 to $x=Q^2/(Q^2+W^2)$. In a diffractive event, a 
colour-neutral cluster of `wee partons' is stripped off the proton. The 
invariant mass of this cluster and the virtual photon is $M$. Hence, $W$ is 
replaced by $M$, and the quark which couples to the photon can be radiated 
by a parton whose fraction of the cluster momentum varies from 1 to 
$\beta=Q^2/(Q^2+M^2)$. 

\begin{figure}[t]
\begin{center}
\vspace*{-.5cm}
\parbox[b]{15cm}{\psfig{width=15cm,file=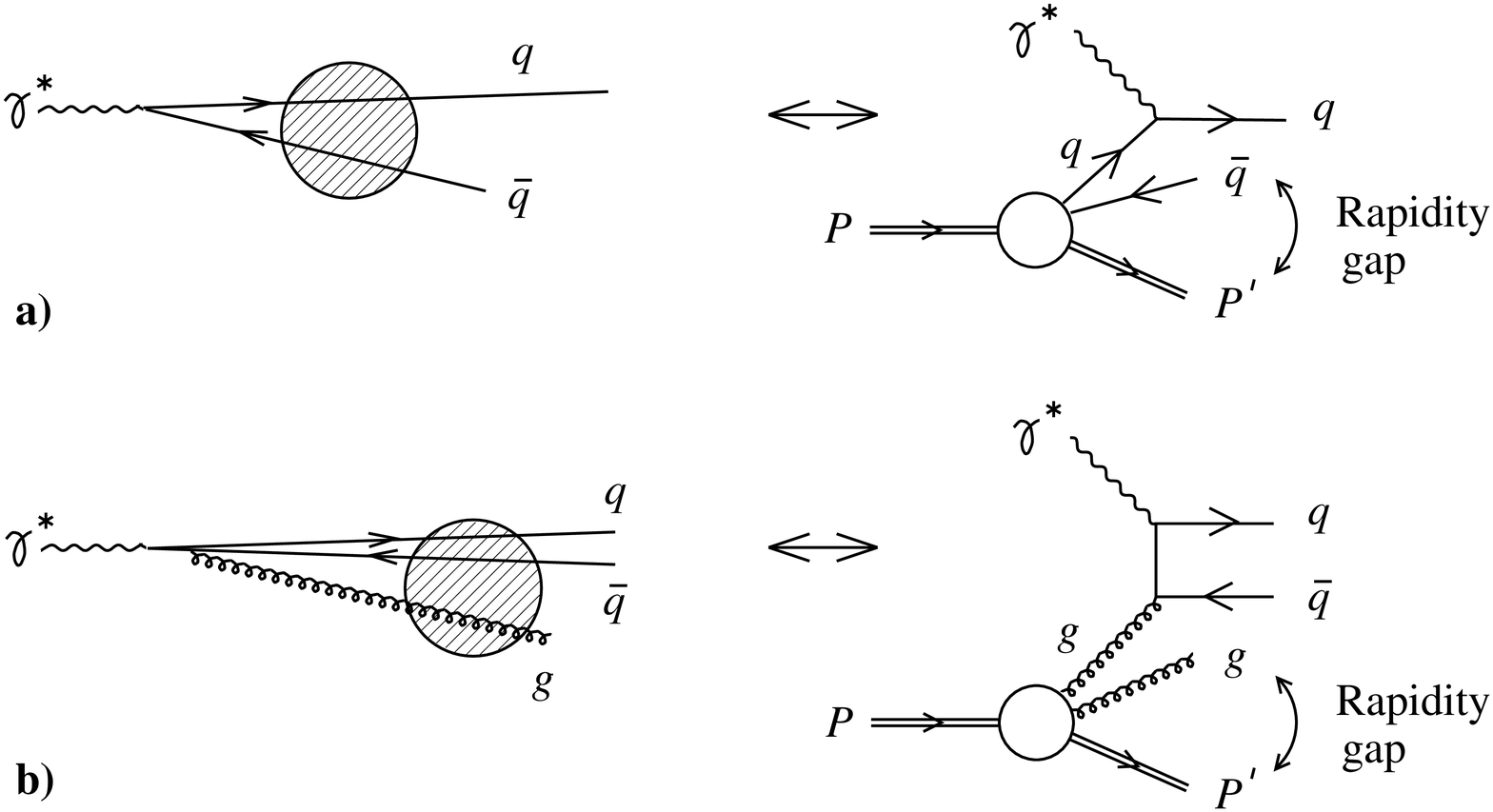}}\\
\end{center}
\refstepcounter{figure}
\label{f2d}
{\bf Figure \ref{f2d}:} Diffractive DIS in the proton rest frame (left) and 
the Breit frame (right); asymmetric quark fluctuations correspond to 
diffractive quark scattering, asymmetric gluon fluctuations to diffractive 
boson-gluon fusion. 
\end{figure}

The separation of the colour-neutral cluster of 
`wee partons' is non-perturbative and independent of the perturbative 
evolution. It is, however, incorrect to visualize a two-step process where 
the colour-neutral cluster is first emitted by the proton and then probed 
by the virtual photon. If this was the case, the two-gluon or two-quark 
cluster relevant in our calculation (cf. the r.h. side of Fig.~\ref{f2d}) 
would necessarily lead to parton distributions symmetric in $\beta$ and 
$1-\beta$. A counter example to this is provided by the model distributions 
derived in Sect.~\ref{av}. 

The diffractive quark and gluon distributions have been determined in 
\cite{h}. In terms of Wilson loops in coordinate space, the quark 
distribution can be expressed as follows (cf. Appendix A), 
\bea
{dq(b,\xi)\over d\xi}&=&{2b\over \xi^2(1-b)^3}
           \int{d^2\k'k'^4\over(2\pi)^6 N_c} 
           \int_{\y,\y'} e^{i\k'(\y-\y')}\,{\y\y'\over y\, y'}\nn\nn
&&\times\,K_1(yN)K_1(y'N)\int_{\x}\mbox{tr}W^{\cal F}_{\x}(\y)
\mbox{tr}W^{\cal F\dagger}_{\x}(\y')\;, \label{dqy} 
\eea
where $N_c$ is the number of colours and 
\be
N^2=k'^2{b\over 1-b}\;.
\ee
The corresponding expression for the diffractive gluon distribution is very 
similar. Because of the different colour and spin of the gluon, the Wilson 
loop is now in the adjoint representation, the tensor structure is different 
and the modified Bessel function $K_2$ appears\footnote{To simplify the 
colour algebra of Sect.~\ref{av}, we use the large $N_c$ limit throughout
this paper, with the exception of the standard DGLAP evolution in 
Sect.~\ref{na}.}, 
\bea
{dg(b,\xi)\over d\xi} &=& {b\over \xi^2(1-b)^3}
\int {d^2 \k'\,k'^4 \over (2\pi)^6 N_c^2}
\int_{\y,\y'} e^{i\k'(\y-\y')}\,t_{ij}(\y)t_{ij}(\y')\nn\nn
&&\times\,K_2(yN)K_2(y'N)\int_{\x}\mbox{tr}W^{\cal A}_{\x}(\y)
\mbox{tr}W^{\cal A\dagger}_{\x}(\y')\;,\label{dgy}
\eea
where
\be
t_{ij}(\y)=\delta_{ij}-2{y_iy_j\over y^2}\;. 
\ee
The physical content of both expressions is rather transparent. The modified 
Bessel functions contain the kinematic effects due to the propagators of the 
partons penetrating the colour field and the Wilson loops represent the 
interaction with the proton. Figure~\ref{f2d} illustrates the correspondence 
between the semiclassical and parton model view of the leading-order 
processes testing diffractive quark and gluon distributions. 

To illustrate the close similarity of diffractive and inclusive scattering
in the semiclassical approach,
it is instructive to evaluate the diffractive contribution to the inclusive 
quark distribution. To achieve this, one sets $x=b\xi$ and integrates 
over $\xi$ keeping $x$ fixed,
\be
q_D(x) = \int_x^1 d\xi {dq(x/\xi,\xi)\over d\xi}\;.
\ee
After neglecting terms ${\cal O}(x)$ and exchanging the integration variable 
$\xi$ for $N^2$, the $k_\perp'$ integration can be performed trivially. 
As a consequence, the integrand is evaluated at $\y=\y'$, giving the simple 
result 
\be
xq_D(x)={2\over (2\pi)^4N_c}\int N^2 dN^2 \int_{\y} K_1(y N)^2 
              \int_{\x} \mbox{tr}W^{\cal F}_{\x}(\y)
              \mbox{tr}W_{\x}^{\cal F\dagger}(\y)\;. \label{qd}
\ee
This expression can also be obtained from Eq.~(\ref{qi}) by the substitution 
\be
\mbox{tr}\left(W^{\cal F}_{\x}(\y)W_{\x}^{{\cal F}\dagger}(\y)\right) 
\rightarrow \frac{1}{N_c}\mbox{tr}W^{\cal F}_{\x}(\y)\mbox{tr}
W_{\x}^{\cal F\dagger}(\y)\,.
\ee
After this substitution, the $N^2$ integration in Eq.~(\ref{qi}) 
becomes UV-finite, so that the upper limit $\mu^2$ can be dropped; the 
$\ln Q^2$ term disappears since 
$\mbox{tr}\,\partial_{\y}W^{\cal F}_{\x}(0)=0$. The above relation between 
the inclusive and diffractive quark distribution illustrates most clearly the 
basic idea of the semiclassical approach: diffractive and non-diffractive 
events are both induced by the interaction of the soft proton colour 
field with a partonic fluctuation of the incoming virtual photon. 
The different final states are realized by projecting the outgoing 
partonic system onto different colour configurations. 

Let us finally note that the semiclassical formulae for diffractive parton 
distributions \cite{h} used in this paper are sufficiently general to 
accommodate different models of the hadron colour field. For example, they 
can serve as an alternative starting point for the derivation of 
the perturbatively generated distributions of \cite{hks}, where a heavy 
quark-antiquark fluctuation of a photon was used as a model for the target. 
Taking the colour field responsible for tr$W$tr$W^\dagger$ 
in Eqs.~(\ref{dqy}) and 
(\ref{dgy}) to be the perturbative field of a small dipole configuration, 
the results of~\cite{hks} are exactly reproduced (cf. Appendix B). However, 
the analysis of the present paper is based on the fundamentally 
non-perturbative model described in the next section.

\section{A Model for the gluon field averaging}\label{av}
The averaging over the gluon field configurations of the target discussed in 
the two previous sections is a complicated operation depending on the full 
details of the non-perturbative hadronic state. However, in the special case 
of a very large target, a quantitative treatment becomes possible under 
minimal additional assumptions.

McLerran and Venugopalan have observed that the large size of a hadronic 
target, realized, e.g., in an extremely heavy nucleus, introduces a new 
hard scale into the process of DIS \cite{mv}. From the target rest frame 
point of view, this means that the typical transverse size of the partonic 
fluctuations of the virtual photon remains perturbative \cite{hw}, thus 
justifying the omission of higher Fock states in the semiclassical 
calculation. Note that this does not imply a complete reduction to 
perturbation theory since the long distance which the partonic fluctuation 
travels in the target compensates for its small transverse size, thus 
requiring the eikonalization of gluon exchange. 

\begin{figure}[ht]
\begin{center}
\vspace*{-.5cm}
\parbox[b]{12cm}{\psfig{width=9cm,file=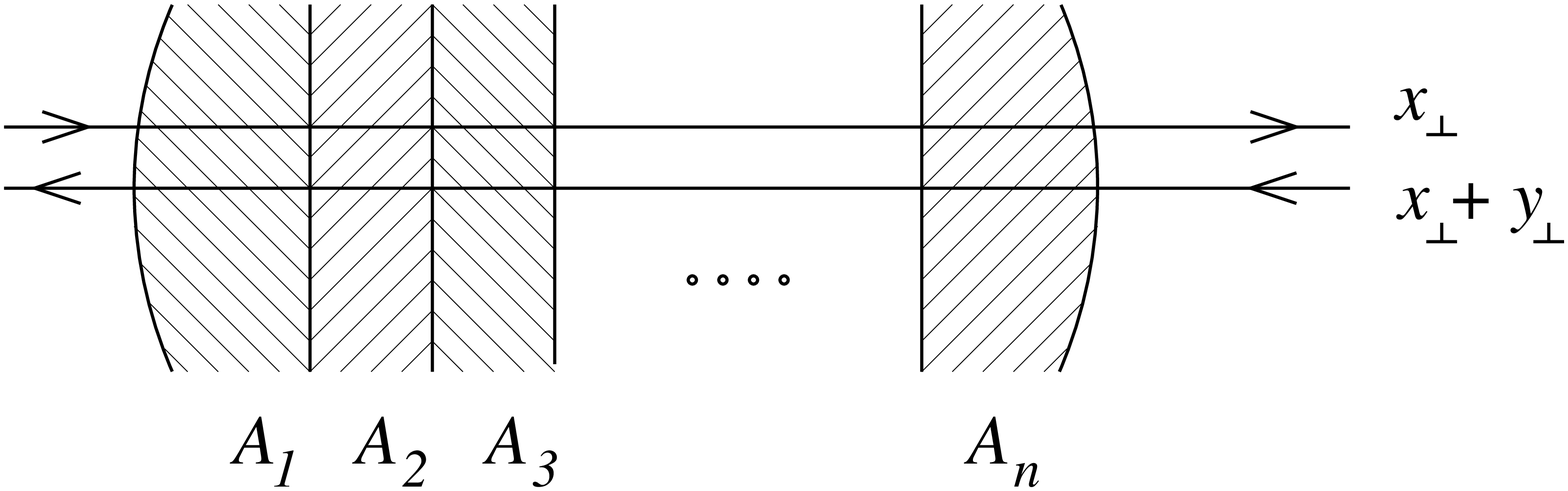}}\\
\end{center}
\refstepcounter{figure}
\label{lt}
{\bf Figure \ref{lt}:} Colour dipole travelling through a large
hadronic target.
\end{figure}

Within this framework, it is natural to introduce the additional assumption 
that the gluonic fields encountered by the partonic probe in distant regions 
of the target are not correlated (cf. \cite{jkmw} and the somewhat 
simplified discussion in \cite{hw}). Thus, one arrives at the situation 
depicted in Fig.~\ref{lt}, where a colour dipole passes a large number of 
regions, each one of size $\sim 1/\Lambda$, with mutually uncorrelated 
colour fields $A_1$ ... $A_n$.

Consider the fundamental quantity $W_{x_\perp}(y_\perp)_{ij}[A]\,
W^\dagger_{x_\perp}(y_\perp')_{kl}[A]$ which, after specifying the 
required representation and appropriately contracting the colour indices 
$ijkl$, enters the formulae for inclusive and diffractive parton 
distributions. According to Eq.~(\ref{wuu}), this quantity is the sum of 
four terms, the most complicated of which involves four $U$ matrices,
\vspace*{.2cm}
\bea
&&\hspace*{-1.3cm}
\left\{U_{x_\perp}[A]\,U^\dagger_{x_\perp+y_\perp}[A]\right\}_{ij}\left\{
U_{x_\perp+y_\perp'}[A]\,U^\dagger_{x_\perp}[A]\right\}_{kl}\label{um}
\\
&&\hspace*{3cm}
=\left\{U_{x_\perp}[A_n]\cdots U_{x_\perp}[A_1]\,\,\,
U^\dagger_{x_\perp+y_\perp}[A_1]\cdots U^\dagger_{x_\perp+y_\perp}[A_n]
\right\}_{ij}
\nonumber
\\
&&\hspace*{3cm}\vspace*{.2cm}
\times\left\{U_{x_\perp+y_\perp'}[A_n]\cdots U_{x_\perp+y_\perp'}[A_1]\,\,\,
U^\dagger_{x_\perp}[A_1]\cdots U^\dagger_{x_\perp}[A_n]\right\}_{kl}\,.
\nonumber
\eea
The crucial assumption that the fields in regions $1$ ... $n$ are 
uncorrelated is implemented by writing the integral over all field 
configurations as
\be
\int_{A}=\int_{A_1}\cdots\int_{A_n}\,\,,\label{int}
\ee
i.e., as a product of independent integrals. Here the appropriate weighting 
provided by the target wave functional is implicit in the symbol $\int_A$. 

Under the integration specified by Eq.~(\ref{int}), the $U$ matrices on the 
r.h. side of Eq.~(\ref{um}) can be rearranged to give the result 
\be
\hspace*{-6.5cm}\int_A
\left\{U_{x_\perp}[A]\,U^\dagger_{x_\perp+y_\perp}[A]\right\}_{ij}\left\{
U_{x_\perp+y_\perp'}[A]\,U^\dagger_{x_\perp}[A]\right\}_{kl}\label{ro}
\ee
\[
\hspace*{2cm}
=\int_{A_1}\cdots\int_{A_n}
\left\{U_{x_\perp}[A_1]\,\,U^\dagger_{x_\perp+y_\perp}[A_1]\cdots 
U_{x_\perp}[A_n]\,\,U^\dagger_{x_\perp+y_\perp}[A_n]\right\}_{ij}
\]
\[\hspace*{4cm}
\times\left\{U_{x_\perp+y_\perp'}[A_n]\,\,U^\dagger_{x_\perp}[A_n]\cdots
U_{x_\perp+y_\perp'}[A_1]\,\,U^\dagger_{x_\perp}[A_1]\right\}_{kl}\,.
\]
To see this, observe that the $A_1$ integration acts on the integrand
$\{U_{x_\perp}[A_1]\,U^\dagger_{x_\perp+y_\perp}[A_1]\}_{i'j'}$ $\{
U_{x_\perp+y_\perp'}[A_1]\,U^\dagger_{x_\perp}[A_1]\}_{k'l'}$ transforming 
it into an invariant colour tensor with the indices $i'j'k'l'$. The 
neighbouring matrices $U_{x_\perp}[A_2]$ and $U^\dagger_{x_\perp}[A_2]$ can 
now be commuted through this tensor structure in such a way that the 
expression $\{U_{x_\perp}[A_2]\,U^\dagger_{x_\perp+y_\perp}[A_2]\}_{i''j''}$ 
$\{U_{x_\perp+y_\perp'}[A_2]\,U^\dagger_{x_\perp}[A_2]\}_{k''l''}$ emerges. 
Subsequently, the $A_2$ integration transforms this expression into an 
invariant tensor with indices $i''j''k''l''$. Repeating this 
argument, one eventually arrives at the structure displayed on the r.h. side 
of Eq.~(\ref{ro}).

To evaluate Eq.~(\ref{ro}) further, observe that it represents a contraction 
of $n$ identical tensors 
\be
F_{ijkl}=\int_{A_m} \{U_{x_\perp}[A_m]\,U^\dagger_{x_\perp+y_\perp}[A_m]\}_{ij}
\,\{U_{x_\perp+y_\perp'}[A_m]\,U^\dagger_{x_\perp}[A_m]\}_{kl}\,,\label{ft}
\ee
where the index $m$ refers to any one of the 
regions $1$ ... $n$ into which the target is subdivided. At this 
point, we make use of the fact that for a sufficiently large target 
the transverse separations $y_\perp$ and $y_\perp'$ are always small 
\cite{hw}. In fact, for a target of geometrical size $\sim n/\Lambda$ 
(where $n\gg 1$), the relevant transverse distances are bounded by 
$y^2\sim y'^2\sim 1/n\Lambda^2$. 

Assuming that size and $\x$ dependence of typical field configurations $A_m$ 
are characterized by the scale $\Lambda$, it follows that the products 
$U_{\x}U^\dagger_{\x+\y}$ and $U_{\x+\y'}U^\dagger_{\x}$ are close to unit
matrices for all relevant $\y$ and $\y'$. Therefore, it is justified to 
write 
\be
U_{x_\perp}[A_m]\,U^\dagger_{x_\perp+y_\perp}[A_m]=\exp\big\{iT^af^a(\x,\y)[A_m]
\big\}\,,\label{exp}
\ee
where $T^a$ are the conventional group generators and $f^a$ are functions 
of $\x$ and $\y$ and functionals of $A_m$. Equation~(\ref{exp}) and its $\y'$ 
analogue are expanded around $\y=\y'=0$ (which corresponds to $f^a(\x,0)=0$) 
and inserted into Eq.~(\ref{ft}). At leading non-trivial order, the result 
reads 
\be
F_{ijkl}=\delta_{ij}\delta_{kl}\left(1-\frac{1}{2}\gamma C_R(y^2+
y'^2)\right)+\gamma(y_\perp y_\perp')T^a_{ij}T^a_{kl}\,,\label{ftf}
\ee
where $C_R$ is the Casimir number of the relevant representation 
($C_R=C_{F,A}$) and $\gamma$ is defined by 
\be
\int_A f^a(\x,\y)f^b(\x,\y') = \gamma \delta^{ab} (\y\y')+{\cal O}(y^2y'^2)
\,.\label{ga}
\ee
Note that the absence of terms linear in $f^a$ and the simple structure on 
the r.h. side of Eq.~(\ref{ga}) are enforced by colour covariance and 
transverse space covariance. The absence of an explicit $\x$ dependence is 
a consequence of the homogeneity that we assume to hold over the large 
transverse size of the target. Neglecting boundary effects, we can account 
for the $\x$ integration by multiplying the final result with a parameter 
$\Omega\sim n^2/\Lambda^2$ that characterizes the geometrical cross section 
of the target. 

Substituting the $n$ tensors $F_{ijkl}$ on the r.h. side of Eq.~(\ref{ro}) 
by the expression given in Eq.~(\ref{ftf}) and contracting the colour 
indices as appropriate for the inclusive and diffractive case respectively, 
one obtains, in the large-$N_c$ limit, 
\be
\int_A\left\{U_{x_\perp}U^\dagger_{x_\perp+y_\perp}\right\}_{ij}\left\{
U_{x_\perp+y_\perp'}U^\dagger_{x_\perp}\right\}_{ji}=d_R\left[1-\frac{1}{2}
\gamma C_R(y_\perp-y_\perp')^2\right]^n\,,\label{uu1}
\ee
\be
\int_A\left\{U_{x_\perp}U^\dagger_{x_\perp+y_\perp}\right\}_{ii}\left\{
U_{x_\perp+y_\perp'}U^\dagger_{x_\perp}\right\}_{jj}=d_R^2\left[1-\frac{1}{2}
\gamma C_R(y_\perp^2+y_\perp'^2)\right]^n\,,\label{uu2}
\ee
where $d_R$ is the dimension of the representation. 

Since $n$ is assumed to be large and the typical values of $y^2$ and $y'^2$ 
do not exceed $1/n\Lambda^2$, the formula $(1-x/n)^n\simeq\exp[-x]$ can be 
applied to the r.h. sides of Eqs.~(\ref{uu1}) and (\ref{uu2}). Furthermore, 
contributions proportional to $\{U_{\x}U^\dagger_{\x+\y}\}_{ij}\delta_{kl}
\,,\quad\delta_{ij}\{U_{\x+\y'}U^\dagger_{\x}\}_{kl}$ and $\delta_{ij}
\delta_{kl}$ have to be added to obtain the complete expression for 
$W_{x_\perp}(y_\perp)_{ij}\,W^\dagger_{x_\perp}(y_\perp')_{kl}$. The 
corresponding calculations are straightforward and the result reads 
\begin{eqnarray}
\int_{x_\perp}\int_A\mbox{tr}\left(W_{x_\perp}(y_\perp)W^\dagger_{x_\perp}
(y_\perp')\right)&=&\Omega d_R\left[1-e^{-a_Ry^2}-e^{-a_Ry'^2}+e^{-a_R(y_\perp-
y_\perp')^2}\right]\,,\label{ww1}
\\
\int_{x_\perp}\int_A\mbox{tr}W_{x_\perp}(y_\perp)\mbox{tr}
W^\dagger_{x_\perp}(y_\perp')&=&\Omega d_R^2\left[1-e^{-a_Ry^2}\right]\,
\left[1-e^{-a_Ry'^2}\right]\,,\label{ww2}
\end{eqnarray}
where $a_R=n\gamma C_R/2$ plays the role of a saturation scale.

The above calculation, performed at large $N_c$ and for the case of a large 
target subdivided into many uncorrelated regions, has no immediate 
application to realistic experiments. However, it provides us with a 
set of non-perturbative inclusive and diffractive 
parton distributions which are highly constrained with respect to each other. 
Assuming that some of the essential features of diffractive and inclusive 
DIS are common to both the above model and the realistic proton case, we 
use the basic formulae Eq.~(\ref{ww1}) and (\ref{ww2}) for a 
phenomenological analysis. For this purpose, $\Omega$ and $a\equiv n\gamma 
N_c/4$ are considered as new fundamental parameters, giving rise to the 
following formulae for the basic hadronic quantities required in 
Sections~\ref{f2sec} and \ref{f2dsec}, 
\begin{eqnarray}
\int_{x_\perp}\int_A\mbox{tr}\left(W^{\cal F}_{x_\perp}(y_\perp)
W^{{\cal F}\dagger}_{x_\perp}(y_\perp')\right)&=&\Omega N_c\left[1-e^{-ay^2}-
e^{-ay'^2}+e^{-a(y_\perp-y_\perp')^2}\right]\label{ww0}\,,
\\
\frac{1}{N_c}\int_{x_\perp}\int_A\mbox{tr}W^{\cal F}_{x_\perp}(y_\perp)
\mbox{tr}W^{{\cal F}\dagger}_{x_\perp}(y_\perp')&=&\Omega N_c\left[1-
e^{-ay^2}\right]\,\left[1-e^{-ay'^2}\right]\,,\label{wwf}
\\
\frac{1}{N_c^2}\int_{x_\perp}\int_A\mbox{tr}W^{\cal A}_{x_\perp}(y_\perp)
\mbox{tr}W^{{\cal A}\dagger}_{x_\perp}(y_\perp')&=&\Omega N_c^2\left[1-
e^{-2ay^2}\right]\,\left[1-e^{-2ay'^2}\right]\,.\label{wwa}
\end{eqnarray}
A similar, Glauber type $y^2$ dependence has been recently used in the DIS 
analysis of \cite{gw}. 
Note that according to Eqs.~(\ref{ww0})--(\ref{wwa}) the 
diffractive structure function is not suppressed by a colour factor relative 
to the inclusive structure function, as originally suggested in~\cite{bh}.

\section{Numerical analysis}\label{na}

The model of the previous section provides an example for the relation between 
diffractive and inclusive parton distributions in the semiclassical 
approach. Although the derivation was based on a large hadronic target 
(with radius much greater than $1/\Lambda$), we expect some qualitative 
features of the resulting distributions to apply to the proton as well. 
The above model distributions are used as non-perturbative input at some 
small scale $Q_0^2$ and are evolved to higher $Q^2$ using the leading-order 
DGLAP equations~\cite{dglap}. The non-perturbative parameters of the 
model, as well as the scale $Q_0^2$, are then determined from a combined 
analysis of experimental data on inclusive and diffractive structure 
functions. 

At first sight, the semiclassical description of parton distribution 
functions always predicts an energy dependence corresponding to a classical 
bremsstrahlung spectrum: $q(x),g(x)\sim 1/x$. However, this na\"\i ve 
prediction assumes the averaging over the soft field configurations inside 
the proton to be independent of the energy of the quark pair used to probe 
these configurations. As already outlined in Sect.~\ref{f2sec}, one expects 
that, in a more complete treatment, a non-trivial energy dependence is 
induced since the averaging procedure encompasses more and more modes of 
the proton field with increasing energy of the probe. We are, however, 
unable to calculate this non-perturbative energy dependence from first 
principles. Instead, we choose to parametrize it in the form of a soft, 
logarithmic growth of the normalization of diffractive and inclusive parton 
distributions with the collision energy $\sim 1/x$, consistent with the 
unitarity bound. This introduces one further parameter, $L$, into the model: 
\begin{equation}
\Omega \to \Omega \left(L - \ln x \right)^2.
\end{equation}

Including this energy dependence, our 
model yields the following compact
expressions for the semiclassical inclusive 
parton distributions Eqs.~(\ref{qi}),(\ref{gi}) at a  low scale $Q_0^2$:
\begin{eqnarray}
xq(x,Q_0^2) & = & \frac{a \Omega N_c \left( L - \ln x \right)^2}
{3 \pi^3} \left(\ln\frac{Q_0^2}{a} - 0.6424\right)\; , \label{qiinp}\\
xg(x,Q_0^2) & = & \frac{ 2 a \Omega N_c \left( L - \ln x \right)^2}
{\pi^2 \alpha_s(Q_0^2)}\; \label{giinp}. 
\end{eqnarray}
Being derived in the semiclassical approach, these expressions are only 
valid in the small-$x$ region, which we define by $x\leq 0.01$. 
In our numerical analysis, we shall multiply the above expressions 
with $(1-x)$ to ensure vanishing of the distributions in the limit $x\to 1$, 
which is required for the numerical stability of the DGLAP evolution.

The corresponding expressions for the diffractive distributions 
Eqs.~(\ref{dqy}),(\ref{dgy}) can be derived in a similar manner. The 
integrations over the momentum variables are  
outlined in Appendix A. One is then left with the expressions
\begin{eqnarray}
\frac{dq\left(\beta,\xi,Q_0^2\right)}{d\xi} & = & \frac{a\Omega N_c 
(1-\beta) 
\left(L - \ln \xi\right)^2}{2\pi^3\xi^2} f_q(\beta)\; , \label{qdinp} \\
\frac{dg\left(\beta,
\xi,Q_0^2\right)}{d\xi} & = & \frac{a\Omega N_c^2 (1-\beta)^2 
\left(L - \ln \xi\right)^2}{2\pi^3 \beta \xi^2} f_g(\beta)\; \label{gdinp},
\end{eqnarray}
with the functions $f_{q,g}(\beta)$ being defined in Appendix A. 
The $\beta$ spectrum of the diffractive
parton distributions at $Q_0^2$ is independent of the unknown 
non-perturbative parameters. Note that our model does not specify whether, 
in the diffractive case, the energy-dependent logarithm  should be a 
function of $x$ or of $\xi$. However, both prescriptions differ only by 
terms proportional to $\ln \beta$, which can be disregarded in comparison 
with $\ln x$ or $\ln \xi$ in the small-$x$ region. Like their inclusive 
counterparts, these distributions are only valid in the region $\xi \leq 
0.01$, where we believe the semiclassical approach to apply. 

The above equations summarize our input distributions, depending on 
$a$, $\Omega$, $L$, and on the scale $Q_0^2$, at which these distributions 
will be used as boundary condition for the leading-order 
DGLAP evolution. At this order, the measured structure function 
$F_2$ coincides with the transverse structure function discussed in 
Sects.~\ref{f2sec} and~\ref{f2dsec}. In 
defining structure functions and parton distributions, we assume 
all three light quark flavours to yield the same contribution, such that the
singlet quark distribution is 
simply  six times the quark distribution defined above, both in the 
inclusive and in the diffractive case:
\begin{equation}
\Sigma (x,Q^2) = 6\, q(x,Q^2)\; , \qquad \frac{d\,\Sigma(\xi,\beta,Q^2)}{d\xi} 
= 6\,  \frac{dq(\xi,\beta,Q^2)}{d\xi}\; .
\end{equation}
Valence quark contributions are  
absent in the semiclassical approach, which does not account for the 
exchange of flavour quantum numbers between the proton and the 
fast moving virtual photon state. 

Charm quarks are treated entirely as massive quarks in the fixed flavour 
number scheme \cite{ffns}, which is appropriate since, in the $Q^2$ range 
under consideration, charm threshold effects are far more 
important than the resummation of $\ln Q^2/m_c^2$ terms. We thus fix $n_f=3$ 
in the DGLAP splitting functions and evolve only gluon and singlet quark 
distribution. The structure functions $F_2$ and $F_2^{D(3)}$ are then given 
by the singlet quark distribution and a massive 
charm quark contribution due to boson-gluon fusion. Explicit formulae 
can, for example, be found in \cite{gs}. For our numerical studies we use 
$\Lambda_{{\rm LO},n_f=3}= 144$~MeV ($\alpha_s(M_Z)=0.118$), $m_c=1.5$~GeV,
$m_b=4.5$~GeV,
and we evaluate the massive charm quark contribution for a renormalization 
and factorization scale $\mu_c = 2 m_c$.  

The resulting structure functions can be compared with HERA data on the 
inclusive structure function $F_2(x,Q^2)$~\cite{incl} and on the 
diffractive structure function $F_2^{D(3)}(\xi,\beta,Q^2)$~\cite{diff}. 
These data sets from the H1 and ZEUS experiments are used to determine 
the unknown parameters of our model. We apply the following selection criteria 
to the data: $x\leq 0.01$ and $\xi \leq 0.01$ 
are needed to justify the semiclassical description of the 
proton colour field; with  $Q_0^2$ being a fit parameter, we demand a 
sufficiently large minimum $Q^2=2$~GeV$^2$ to avoid that the data selection 
is influenced by the current value of $Q_0^2$; finally
we require   $M^2 > 4$~GeV$^2$ in the diffractive case to 
justify the leading-twist analysis.  

\begin{figure}[t]
\begin{center}
\parbox[b]{15.5cm}{\psfig{file=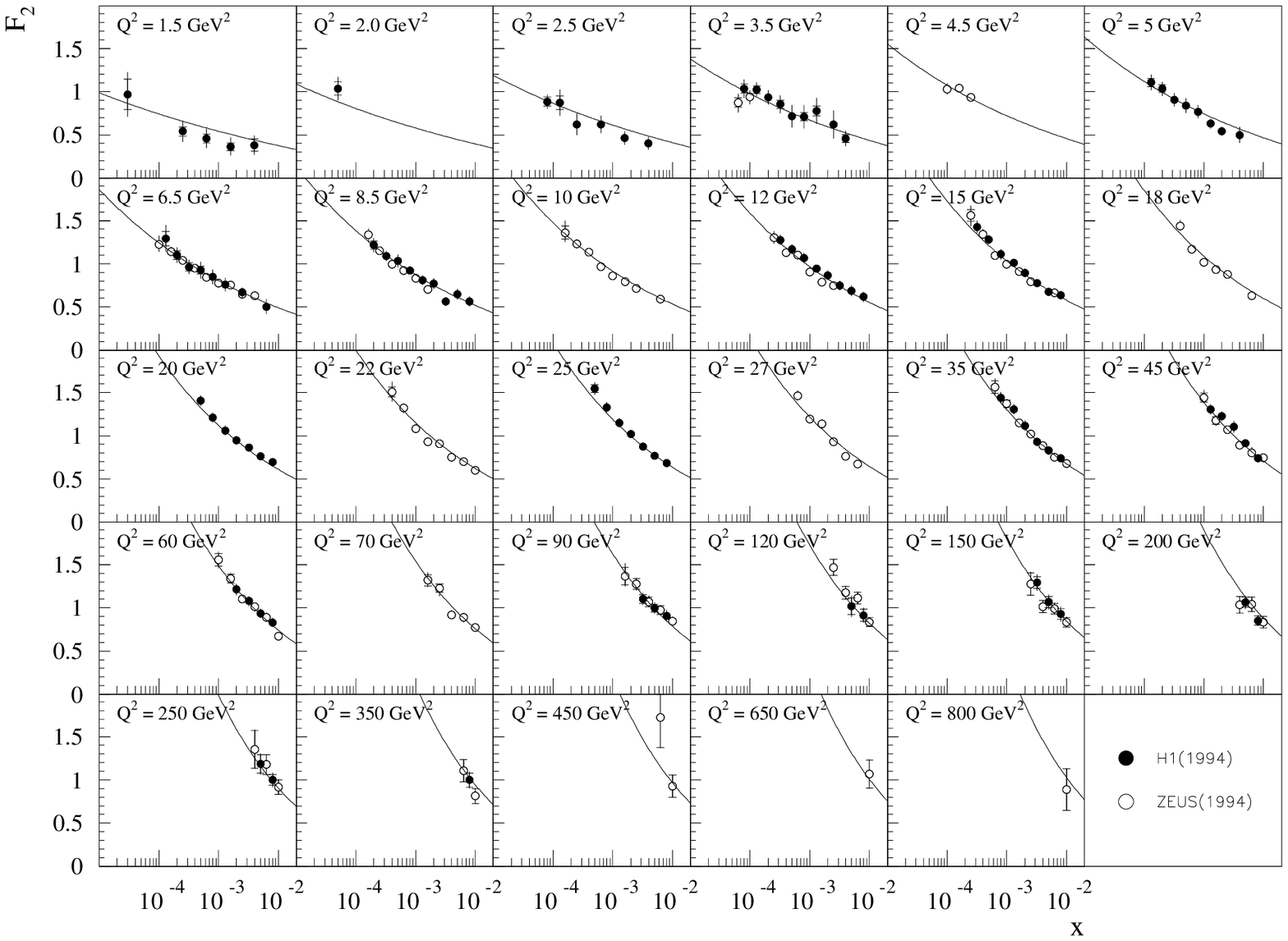,width=15.5cm}}\\
\end{center}
\refstepcounter{figure}
\label{figf2i}
{\bf Figure \ref{figf2i}:}
The inclusive structure function $F_2(x,Q^2)$ at small $x$ 
computed in the semiclassical approach, using the fitted parameters 
given in the text.
Data taken from~\protect\cite{incl}. The data with $Q^2 = 1.5$~GeV$^2$ 
are not included in the fit.
\end{figure}

\begin{figure}[t]
\begin{center}
\parbox[b]{14cm}{\psfig{file=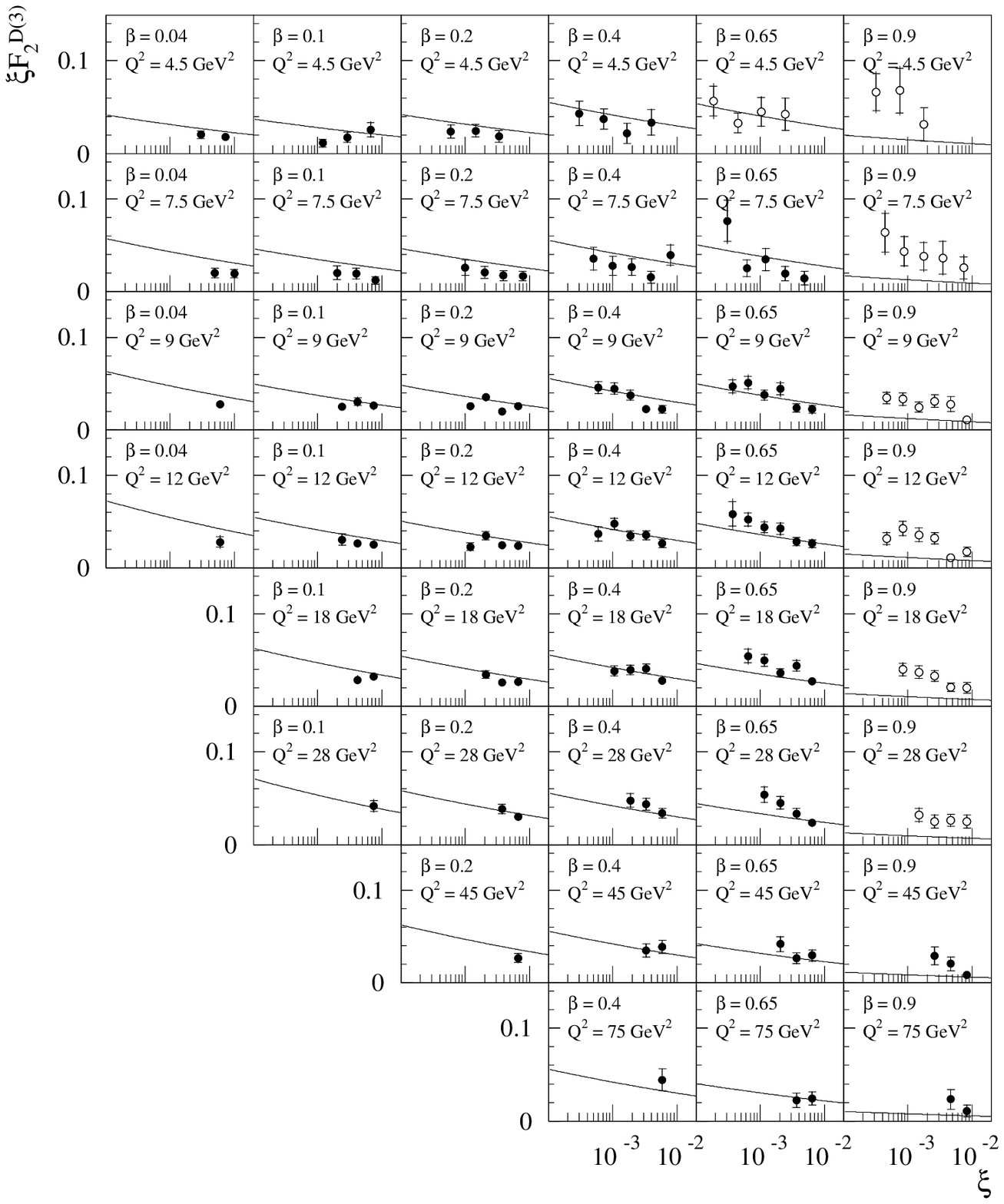,width=14cm}}\\
\end{center}
\refstepcounter{figure}
\label{figf2dh1}
{\bf Figure \ref{figf2dh1}:}
The diffractive structure function $F_2^{D(3)}(\xi,\beta,Q^2)$ at small $\xi$ 
computed in the semiclassical approach, using the fitted parameters 
given in the text.
H1 data taken from~\protect\cite{diff}. The open data points correspond
to $M^2 \leq 4$~GeV$^2$ and are not included in the fit.
\end{figure}

\begin{figure}[t]
\begin{center}
\parbox[b]{10cm}{\psfig{file=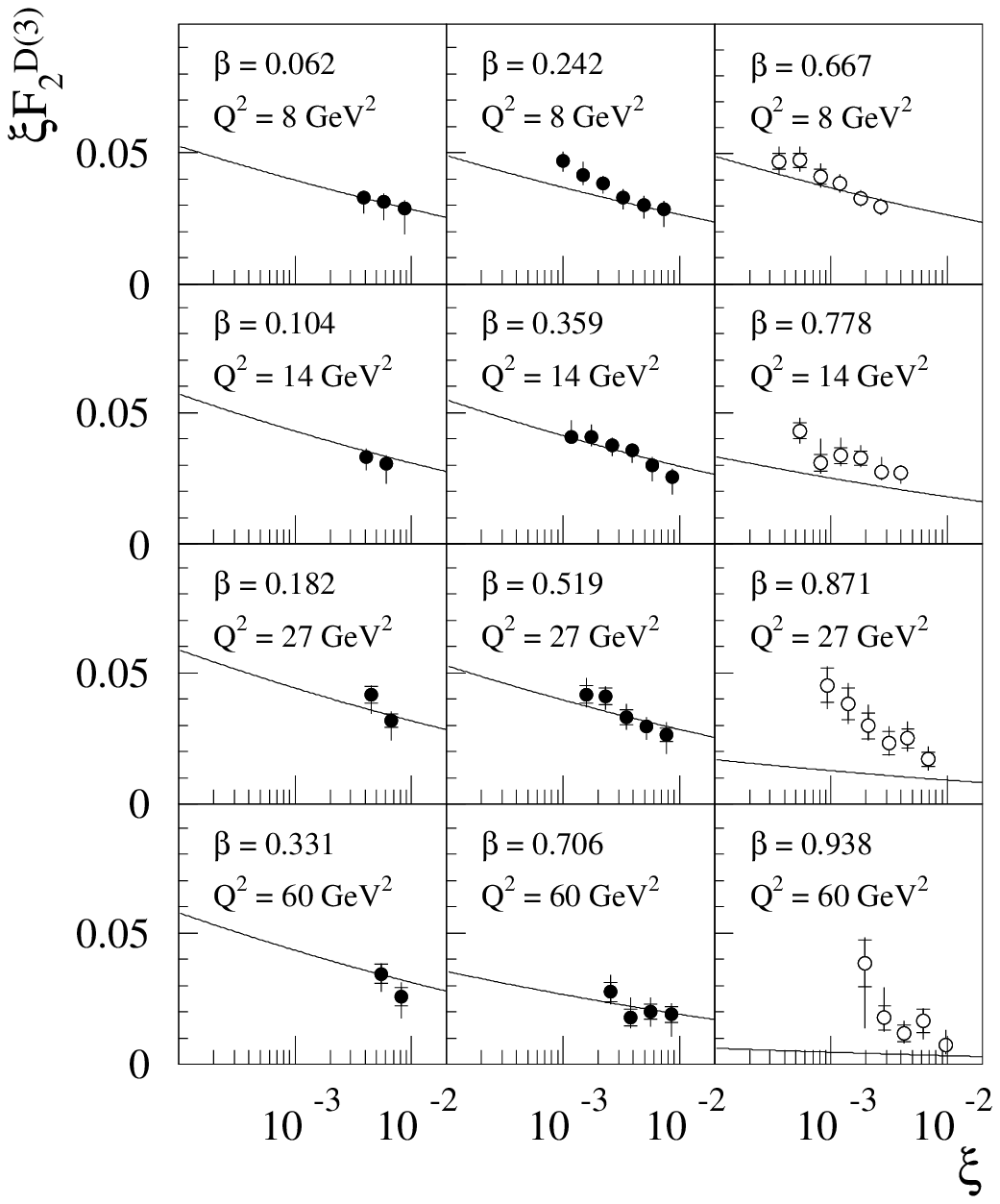,width=10cm}}\\
\end{center}
\refstepcounter{figure}
\label{figf2dzeus}
{\bf Figure \ref{figf2dzeus}:}
The diffractive structure function $F_2^{D(3)}(\xi,\beta,Q^2)$ at small $\xi$ 
computed in the semiclassical approach, using the fitted parameters 
given in the text.
ZEUS data taken from~\protect\cite{diff}. The open data points correspond
to $M^2 \leq 4$~GeV$^2$ and are not included in the fit.
\end{figure}

We determine the optimum set of model parameters from a minimization of the 
total $\chi^2$ (based on 
statistical errors only) 
of the selected data, using the MINUIT  
package~\cite{minuit}. The resulting set of parameters is 
\begin{eqnarray}
Q_0^2 & =& 1.23 \; {\rm GeV}^2\; , \nonumber\\
L &=& 8.16 \; , \nonumber\\
\Omega &=& (712\;   {\rm MeV})^{-2} \; , \nonumber\\
a & = & \left( 74.5\;  {\rm MeV} \right)^2 \; . \label{fitpar}
\end{eqnarray}
The distributions obtained with these fitted parameters
yield a good qualitative description of all data on 
inclusive and diffractive DIS at small $x$, as illustrated in 
Figs.~\ref{figf2i},~\ref{figf2dh1} and~\ref{figf2dzeus}. All parameters are 
given with a precision which allows to reproduce the plots, but which is 
inappropriate with respect to the crudeness of the model. The starting 
scale $Q_0^2$ is in the region where one would expect the transition between 
perturbative and non-perturbative dynamics to take place; the two 
other dimensionful parameters $\Omega L^2$ and $a$ are both of the order 
of typical hadronic scales. 

Our approach fails to reproduce the data on $F_2^{D(3)}$ for low $M^2$ 
(open dots in Figs.~\ref{figf2dh1} and~\ref{figf2dzeus}). This might 
indicate the importance of higher twist contributions in this region, as 
suggested in~\cite{bekw}. It is interesting to note that a breakdown of the 
leading twist description is also observed for inclusive structure 
functions~\cite{mrst2}, where it occurs for 
similar invariant hadronic masses, namely $W^2\lapprox 4$~GeV$^2$. 

The perturbative evolution of inclusive and diffractive structure functions 
is driven by the gluon distribution, which is considerably larger 
than the singlet quark distribution in both cases. 
The ratio of the inclusive singlet quark and gluon distributions can be 
read off from Eqs.~(\ref{qiinp}) and (\ref{giinp}). With the fit 
parameters obtained above, it turns out that the inclusive gluon 
distribution is about twice as large as the singlet quark distribution.
In contrast, the relative magnitude and the $\beta$ dependence of the 
diffractive distributions are completely independent of the model 
parameters. Moreover, their absolute normalization is, up to the slowly 
varying factor $1/\alpha_s(Q_0^2)$, closely tied to the normalization of the 
inclusive gluon distribution. 

Figure~\ref{figdifpdf} displays the diffractive  distributions 
(multiplied by $\beta$ and thus reflecting the 
distribution of momentum carried by the partons) for 
fixed $\xi=0.003$ and different values of $Q^2$. The $\beta$ dependences 
of the quark and the gluon distribution at $Q_0^2$ 
are substantially different: the quark distribution 
$\beta d\Sigma/d \xi$ is peaked around  
$\beta \approx 0.65 $, thus being harder than the distribution
$\beta (1-\beta)$ suggested in~\cite{dl}. 
It vanishes like $\beta$ for $\beta\to 0$ and 
like $(1-\beta)$ at large $\beta$; the gluon distribution $\beta 
d g/d \xi$, on the other hand, approaches a constant for 
$\beta \to 0$ and falls off like $(1-\beta)^2$ at large $\beta$. 
This asymptotic behaviour in the small- and large-$\beta$ region is in 
agreement with the results obtained in the perturbative approach of 
\cite{hks}. In spite of the $(1-\beta)^2$ behaviour, gluons remain important 
even at large $\beta$, simply due to the large total normalization of this 
distribution (the $\beta$ integral over $\beta dg/d\xi$ at $Q_0^2$ is 
approximately three times the $\beta$ integral over $\beta d\Sigma/d\xi$). 
As a result, the quark distribution does not change with increasing $Q^2$ 
for $\beta\approx 0.5$ and is only slowly decreasing for larger values 
of $\beta$. 

\begin{figure}[t]
\begin{center}
\parbox[c]{6cm}{\psfig{file=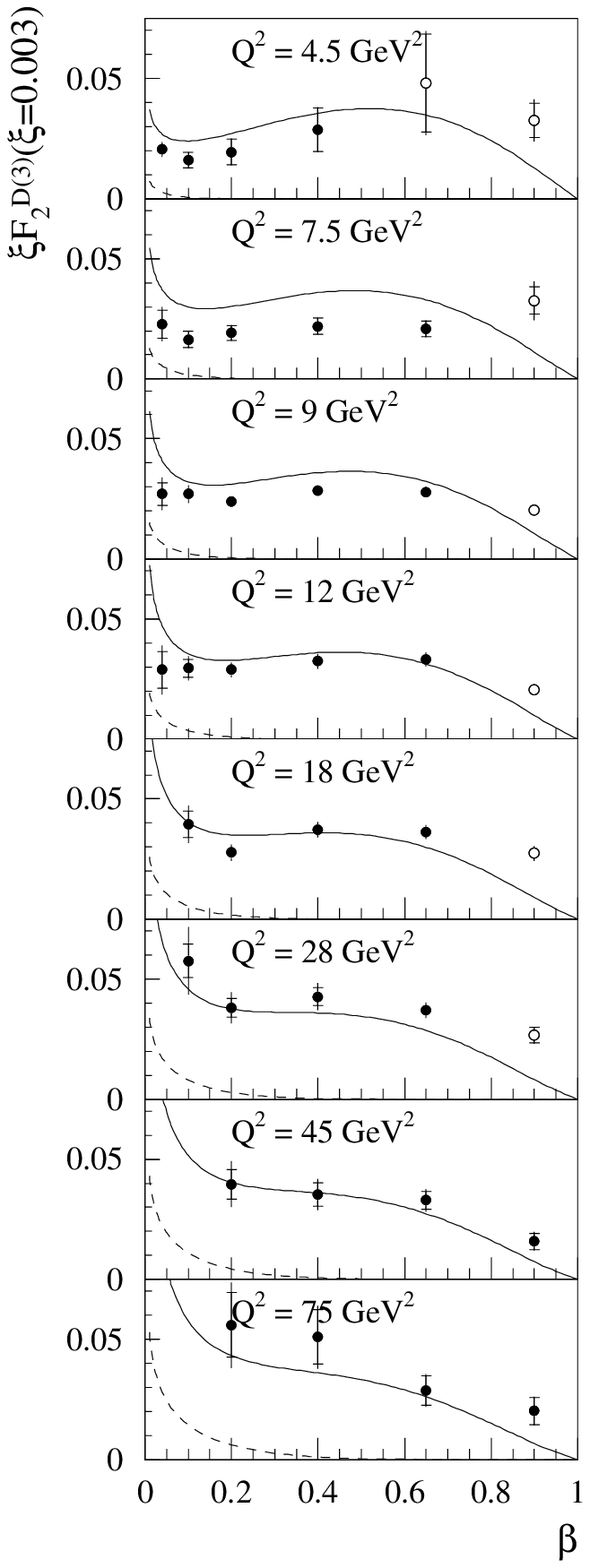,width=6cm} }\parbox[c]{2cm}{\hspace{2cm}}
\parbox[c]{6cm}{\psfig{file=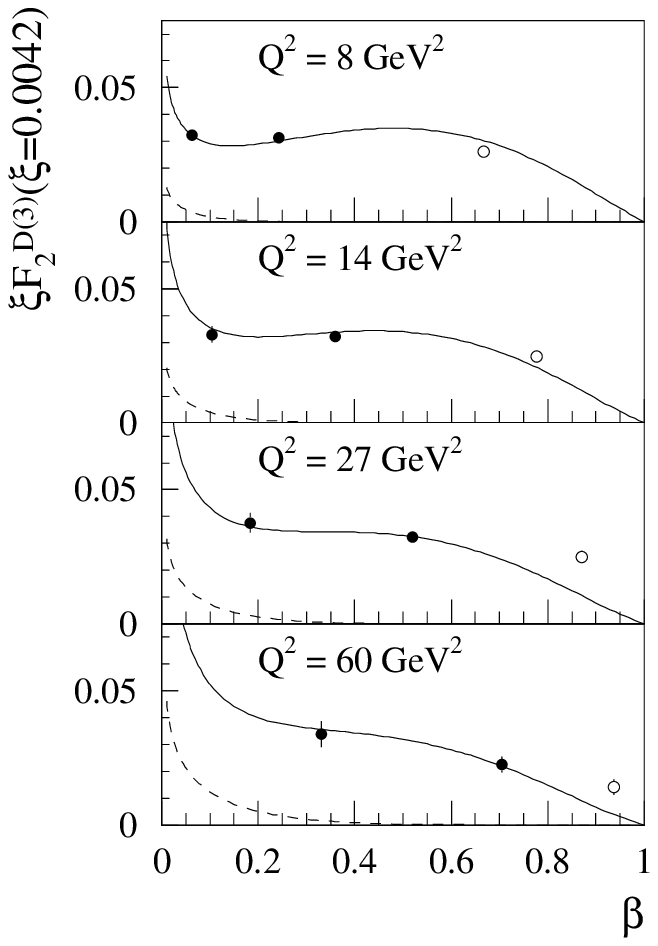,width=6cm} }\\
\vspace{0.6cm}
\end{center}
\refstepcounter{figure}
\label{figdifq2}
{\bf Figure \ref{figdifq2}:}
Dependence of the diffractive structure function $F_2^{D(3)}$ on $\beta$
and $Q^2$, compared to data from H1 (left) and ZEUS 
(right)~\protect\cite{diff}. Open data points 
correspond to $M^2\leq 4$~GeV$^2$. The charm content of the 
structure function is indicated as a dashed line.
\end{figure}

The dependence of the diffractive structure function on $\beta$ and $Q^2$ is 
illustrated in Fig.~\ref{figdifq2}, where we compare our predictions with 
data from the H1 and ZEUS experiments~\cite{diff} at fixed
$\xi = 0.003$ (H1) and $\xi = 0.0042$ (ZEUS). It must be 
pointed out that these data points are based on a combination of data taken 
at various values of $\xi$ (including $\xi > 0.01$), 
which have been extrapolated to fixed $\xi$.
The energy dependence of the diffractive structure function used for 
this extrapolation is different from the energy dependence employed in our 
model, so that a detailed comparison of the data with our results should 
only be made with some caution. Disregarding the large-$\beta$ region, our 
model gives a good description of the $\beta$ dependence of the diffractive 
structure function for all values of $Q^2$. 

The validity of our approach can be tested by studying DIS observables other 
than the diffractive structure 
function $F_2^{D(3)}$ used to fit our model parameters. 
In particular, it would be of interest to study quantities which are 
directly proportional to the diffractive gluon distribution, such as
the charm content of  $F_2^{D(3)}$. The dashed lines in Fig.~\ref{figdifq2} 
show our prediction for the diffractive charm structure function due to 
photon-gluon fusion. Moreover, this structure function is predicted to have 
the same, non-perturbative $\xi$ dependence as $F_2^{D(3)}(\xi,\beta,Q^2)$. 
The charm contribution is sizeable in the small-$\beta$ region. Since the 
full diffractive cross section at fixed $x$ is obtained after integration 
with the measure $d\beta/\beta$, this region yields 
a substantial diffractive charm cross section. 

The $\beta$ dependence of diffractive parton distributions used in the 
present numerical analysis is a direct result of our specific model for the 
colour field averaging. It would therefore be very interesting to repeat the 
analysis with more sophisticated models, such as the stochastic vacuum model 
that was utilized for diffractive meson production in \cite{dnr}.

\section{Conclusions}\label{conc}
In the target rest frame, DIS at small $x$
can be viewed as the interaction of 
a partonic fluctuation of the virtual photon with the proton colour field. 
The semiclassical approach assumes this proton field to be dominated by soft 
modes, which have the sole effect of introducing a non-Abelian phase factor 
for each parton. 
In this approach, a very similar description of inclusive 
and diffractive events emerges, the latter being realized if the partonic 
fluctuation leaves the proton in a colour singlet state. 
Inclusive and diffractive 
structure functions are therefore calculated by evaluating different colour 
contractions of the same soft scattering amplitudes.

Matching the semiclassical and partonic description of structure functions 
at some low scale $Q_0^2$, where logarithmic corrections are still small, 
a set of semiclassical parton distribution functions for inclusive and 
diffractive processes is defined. The evolution to higher scales is 
then determined by the conventional leading-order DGLAP equations.
Initial quark and 
gluon distributions are expressed in terms of averages over the proton 
colour field. 

In the semiclassical framework, a very special role is played by the 
inclusive gluon distribution. In contrast to both the inclusive quark 
distribution and the diffractive quark and gluon distributions, it is only 
sensitive to the short distance structure of the proton field, and it is 
enhanced by an explicit factor $1/\alpha_s$. As a result, the observed 
dominance of the inclusive over the diffractive DIS cross section emerges. 

To study the semiclassical distributions in more detail, we introduce
a non-perturbative model for the proton colour field. This model is derived 
for a very large hadronic target, which can be subdivided into different 
zones of uncorrelated colour field strengths. A non-trivial example of a 
non-perturbative set of semiclassical distributions results. In spite of its 
lacking theoretical justification in the case of a proton target, this set 
of distributions can serve as a basis of a phenomenological analysis. 
Our model depends on two free parameters which are related to the average 
field strength in a zone and to the total geometrical size of the 
target. A further parameter has to be introduced to account for the 
unknown energy dependence induced by the averaging over the field 
configurations. Thus, our model 
describes all semiclassical parton distributions
 by three unknown non-perturbative parameters and 
the matching scale $Q_0^2$.

These three model parameters and $Q_0^2$ are
determined from a combined fit to
measurements of inclusive~\cite{incl} and diffractive~\cite{diff} 
structure functions at small $x$. 
The resulting distributions yield a 
satisfactory description of the structure functions $F_2(x,Q^2)$ and 
$F_2^{D(3)}(\xi,\beta,Q^2)$. 
It turns out that the $Q^2$ evolution of both 
structure functions is mainly driven by large gluon distributions. 

The good agreement of our model with the experimental data allows us 
to conclude that both inclusive and diffractive DIS at small $x$ 
can be described in a unified picture in the semiclassical approach.
A simple model for the colour field averaging, adopted from a large target, 
is used to compute the semiclassical parton distributions at some initial 
scale $Q_0^2$. The behaviour above $Q_0^2$ is then determined by 
perturbative evolution at leading order in $\alpha_s$ and leading twist. 

The rise of $F_2(x,Q^2)$ and of $F_2^{D(3)}(\xi,\beta,Q^2)$ at small $x$ 
have the same, {\it non-perturbative} origin in the energy dependence of 
the average over soft field configurations in the proton. With increasing 
$Q^2$, this rise is enhanced by perturbative evolution in the case of 
the inclusive structure function, while it remains unchanged in the 
diffractive structure function. 

The obtained parton distributions can be used to predict a broad spectrum of 
observables in diffractive and inclusive processes using standard methods of 
perturbative QCD. The qualitative relation between diffractive and inclusive 
DIS and the universal $\xi$ dependence of large-mass diffraction are genuine 
predictions of the semiclassical approach. By contrast, the $\beta$ dependence 
of the diffractive distributions is a result of our specific model for the 
colour 
field averaging. Crucial tests of this model can be performed in future 
measurements of diffractive final states. At the same time, other models for 
the averaging procedure can be tested in the present framework. 
\vspace{.4cm} 

We would like to thank M.F. McDermott, B.R. Webber and H. Weigert for 
valuable discussions and comments.

\section*{Appendix A: Diffractive parton distributions}

In the following, we collect several formulae which are useful in 
connection with the semiclassical expressions for the diffractive 
parton distributions (Sect.~\ref{f2dsec}) and with their explicit 
evaluation within our model of the proton colour field (Sect.~\ref{av}). 

In Ref. \cite{h}, the diffractive quark distribution was given in the form 
\be
{dq(b,\xi)\over d\xi}={2\over \xi^2 (1-b)^2}
\int{d^2\k'\,k'^2\over(2\pi)^8N_c}\int_{\x}\left|\int{d^2k_\perp\,\k
\over u k'^2 + k^2} \int_{\y}e^{i(\k'-\k)\y} \mbox{tr}W^{\cal F}_{\x}
(\y)\right|^2\;,\label{dq}
\ee
where $u=b/(1-b)$. Starting from this result, Eq.~(\ref{dqy}) of 
Sect.~\ref{f2dsec} can be derived using the identity 
\be
\int {d^2\k\over (2\pi)^2} {k_i e^{i\k\y}\over N^2 + \k^2} = {i\over 2\pi} 
{y_i\over y} N K_1(yN)\;.
\ee

For the phenomenological analysis, we use the specific model of 
Sect.~\ref{av}. Inserting for $\mbox{tr} W^{\cal F}_{\x}(\y)\mbox{tr}
W^{{\cal F}\dagger}_{\x}(\y')$ in Eq.~(\ref{dq}) the expression from 
Eq.~(\ref{wwf}), the $\y$ integration and some of the momentum integrations 
can be carried out. The result reads 
\be
{dq(b,\xi)\over d\xi}={a\Omega N_c(1-b)\over 2\pi^3\xi^2} f_q(b)\;, 
\ee
where $f_q(b)$ is an integral over two Feynman-type parameters,
\be
f_q(b) = 4 \int_0^\infty dxdx'\frac{
\left(\frac{\displaystyle \sqrt{b}+x}{\displaystyle (1-b+(\sqrt{b}+x)^2)^2}
\right)\,
\left(\frac{\displaystyle \sqrt{b}+x'}{\displaystyle (1-b+(\sqrt{b}+x')^2)^2}
\right)}
{(x+x')\sqrt{b} + (1-b)\left({\displaystyle x\over \displaystyle \sqrt{b}+x}
+{\displaystyle x'\over \displaystyle \sqrt{b}+x'}\right)}\;.
\ee
We were not able to obtain an analytical expression for this integral. It 
is, however, easily evaluated at $b=0$ and $b=1$ yielding $f_q(0)=1/2$ 
and $f_q(1)=3\pi^2/8-2$.

\begin{figure}[t]
\begin{center}
\parbox[b]{7cm}{\psfig{file=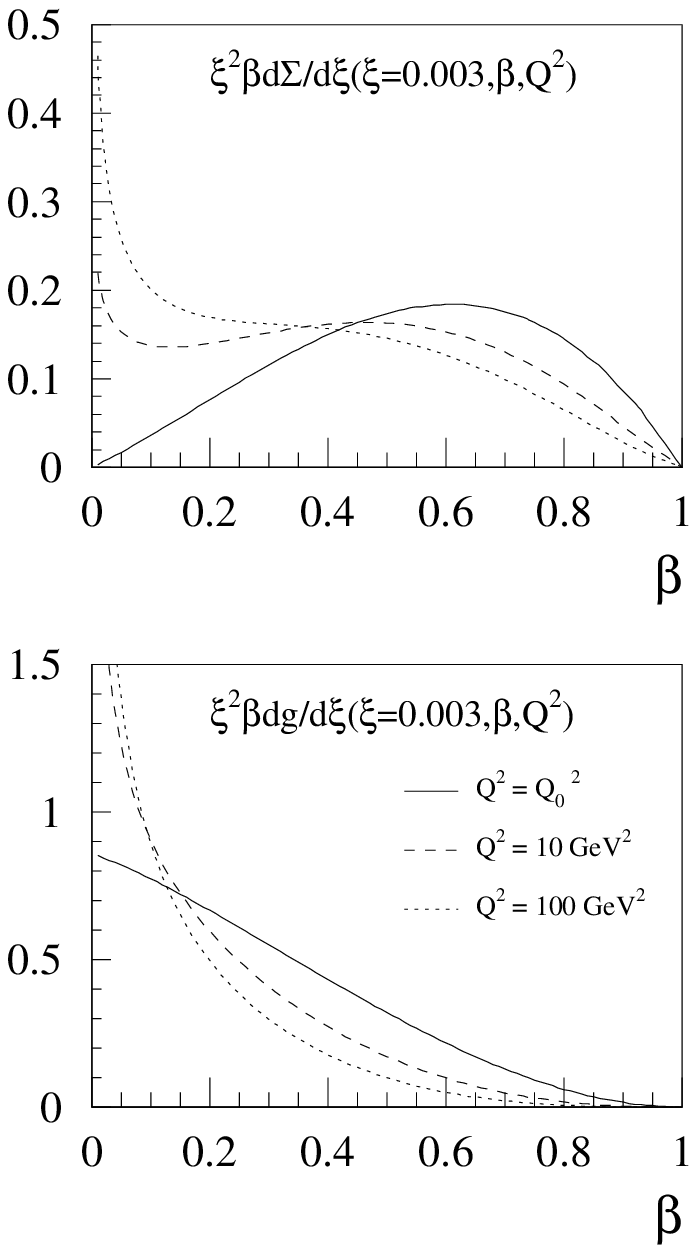,width=7cm}}\\
\end{center}
\refstepcounter{figure}
\label{figdifpdf}
{\bf Figure \ref{figdifpdf}:}
Diffractive quark and gluon distributions at the initial scale $Q_0^2$ and 
after $Q^2$ evolution.
\end{figure}

Similar formulae hold for the diffractive gluon distribution. The result of 
\cite{h} reads
\be
{dg(b,\xi)\over d\xi} = {b\over \xi^2(1-b)^3} \int {d^2 \k'\,k'^4 \over 
(2\pi)^8 N_c^2} \int_{\x}\left|\int {d^2k_\perp\,t_{ij}
\over uk'^2 +k^2 }\int_{\y}e^{i(\k'-\k)\y}\mbox{tr}W^{\cal A}_{\x}(\y)
\right|^2\; ,\label{dg}
\ee
where
\be
t_{ij}=\delta_{ij} + 2\frac{k_i k_j}{uk'^2}\,.
\ee
Using
\be
\int {d^2\k\over (2\pi)^2} \left(\delta_{ij}+2{k_ik_j\over N^2}\right) 
{e^{i\k\y}\over N^2 + \k^2} = {1\over 2\pi} 
\left(\delta_{ij}-2{y_iy_j\over y^2}\right) K_2(yN)\;,
\ee
one obtains Eq.~(\ref{dgy}) in Sect.~\ref{f2dsec}. Inserting for $\mbox{tr}
W^{\cal A}_{\x}(\y)\mbox{tr}W^{{\cal A}\dagger}_{\x}(\y')$ the expression 
in Eq.~(\ref{wwa}) of Sect.~\ref{av}, the $\y$ integration and some of the 
momentum integrations can be carried out. The result has the same structure 
as the diffractive quark distribution, 
\be
{dg(b,\xi)\over d\xi}={a\Omega N_c^2(1-b)^2\over 2\pi^3 \xi^2 b} f_g(b)\;, 
\ee
where $f_g(b)$ is given by the two-dimensional integral
\be
f_g(b) = 2\int_0^\infty dxdx'\frac{
\left(\frac{\displaystyle 1-b+3(1+x)^2b}{\displaystyle (1+x)^2(1-b+(1+x)^2b
)^2}\right)
\left(\frac{\displaystyle 1-b+3(1+x')^2b}{\displaystyle (1+x')^2
(1-b+(1+x')^2b)^2}\right)}
{(x+x')b + (1-b)\left({\displaystyle x\over\displaystyle 1+x}+{
\displaystyle x'\over\displaystyle 1+x'}\right)}\;.
\ee
This integral is easily evaluated for
 $b=0$ and $b=1$ yielding $f_g(0)=4\ln 2$ 
and $f_g(1)=45\pi^2/32-17/2$. 
For general $b$, we have evaluated $f_q(b)$ and $f_g(b)$
only numerically, the results can be inferred from the solid curves in 
Fig.~\ref{figdifpdf}.

\section*{Appendix B: Comparison with a perturbative model}

It is the purpose of this appendix to outline how the perturbative results 
of \cite{hks} can be derived on the basis of the semiclassical formulae for 
diffractive parton distributions of \cite{h} (cf. Eqs.~(\ref{dq}) and 
(\ref{dg}) of this paper). 

The authors of \cite{hks} study diffraction as quasi-elastic scattering off a
special target photon that couples to only one flavour of 
very massive ($M \gg \Lambda$) quarks.
The large quark mass justifies a completely perturbative treatment of 
the target and the diffractive system. In 
this situation, the required $t$ channel colour singlet exchange is realized 
by two gluons coupling to the massive quark loop of the target. In the 
semiclassical approach, these two gluons are understood to be radiated by 
the massive quark loop and are treated as the colour field generating 
tr$W$tr$W^\dagger$. The semiclassical calculation proceeds as follows. 

Equations (\ref{dq}) and (\ref{dg}) have the structure 
\be
\frac{df_a}{d\xi}=F_a\left[\int_{\x}\mbox{tr}W_{\x}\,\mbox{tr}W_{\x}^\dagger
\right]\,,
\ee
where $F_a$ (with $a=q,g$) is a linear functional depending on $\int\mbox{tr}
W_{\x}(\y)\,\mbox{tr}W_{\x}^\dagger(\y')$, interpreted as a function of $\y$ 
and $\y'$. To be differential in $t$, one simply writes 
\be
\frac{df_a}{d\xi\,dt}=\frac{1}{4\pi}F_a\left[\int_{\x}\int_{\x'}\mbox{tr}
W_{\x}\,\mbox{tr}W_{\x'}^\dagger e^{iq_\perp(\x'-\x)}\right]\,,
\ee
with $q_\perp^2=-t$.

The field responsible for tr$W$ is created by a small colour dipole which, 
in turn, is created by the special photon that models the target. At leading 
order in perturbation theory, the colour field of a static quark is 
analogous to its electrostatic Coulomb field. The field of a quark 
travelling on the light cone in $x_-$ direction ($x_{\pm}=x_0\pm x_3$) 
at transverse position $0_\perp$ has therefore the following line integral 
along the $x_+$ direction, 
\be
-\frac{ig}{2}\int A_-\,dx_+\,=\,-ig^2\int\frac{d^2k_\perp}{(2\pi)^2}\,\cdot
\,\frac{e^{ik_\perp\x}}{k_\perp^2}\,\,\,.
\ee
It is exactly this type of line integral that appears in the exponents of 
the non-Abelian phase factors $U$ and $U^\dagger$ that form $W$ 
(cf.~Eq.~(3) of~\cite{h}). A 
straightforward calculation shows that the function tr$W$ produced by a 
dipole consisting of a quark at $\rho_\perp$ and an antiquark at $0_\perp$ 
reads
\bea
\mbox{tr}W_{\x}(\y)&=&-\frac{g^4(N_c^2-1)T_R}{2}\left[\int_{k_\perp,
k_\perp'}\frac{\left(1-e^{-ik_\perp\rho_\perp}\right)\left(1-e^{-ik_\perp'
\rho_\perp}\right)}{(2\pi)^4 k_\perp^2 k_\perp'^2}\right]
\nn
\nn
&&\times \left(1-e^{ik_\perp\y}\right)\left(1-e^{ik_\perp'\y}\right)\,
e^{i(k_\perp+k_\perp')\x}\,,\label{trw}
\eea
where $T_F=1/2$ and $T_A=N_c$ have to be used for the fundamental and 
adjoint representation respectively. 

The final formulae for the diffractive 
parton distributions of the target are obtained after 
integrating over the transverse sizes of the colour dipoles with a weight 
given by the $q\bar{q}$ wave functions of the incoming and outgoing target 
photon. They read
\bea
\frac{df_a}{d\xi\,dt}&\!\!=\!\!&\int dz\,d^2\rho_\perp \int dz'\,d^2
\rho_\perp'\,\frac{1}{4\pi}F_a\left[\int_{\x}\int_{\x'}\mbox{tr}W_{\x}\,
\mbox{tr}W_{\x'}^\dagger e^{iq_\perp(\x'-\x)}\right]\label{faf}
\\
\nn
&&\times \frac{1}{2}\sum_{\epsilon,\epsilon'}\left[\psi_\gamma^*(z,
\rho_\perp,p_\perp',\epsilon_\perp')\,\psi_\gamma(z,\rho_\perp,0_\perp,
\epsilon_\perp)\right]\left[\psi_\gamma^*(z,'\rho_\perp',p_\perp',
\epsilon_\perp')\,\psi_\gamma(z',\rho_\perp',0_\perp,\epsilon_\perp)
\right]\,,\nonumber
\eea
where tr$W_{\x}(\y)$ is produced by the field of a quark at $\rho_\perp$ and 
an antiquark at $0_\perp$, and tr$W_{\x'}(\y')$ is produced by the field of 
a quark at $\rho_\perp'$ and an antiquark at $0_\perp$, as detailed in 
Eq.~(\ref{trw}).

The wave function $\psi_\gamma(z,\rho_\perp,0_\perp,\epsilon_\perp)$ 
characterizes the amplitude for the fluctuation of the incoming target 
photon with polarization $\epsilon$ and transverse momentum $0_\perp$ into 
a $q\bar{q}$ pair with momentum fractions $z$ and $1-z$ and transverse 
separation $\rho_\perp$. Similarly, the wave function $\psi_\gamma^*(z,
\rho_\perp,p_\perp',\epsilon'_\perp)$ characterizes the amplitude for the 
recombination of this $q\bar{q}$ pair into a photon with polarization 
$\epsilon'$ and transverse momentum $p_\perp'=-q_\perp$. The summation over 
the helicities of the intermediate quark states, which are conserved by the 
high-energy gluonic interaction, is implicit. 

The required product of photon wave functions
can be calculated following the lines of~\cite{bhm,w}. It reads explicitly
\vspace*{.3cm}
\be
\hspace*{-9cm}\psi_\gamma^*(z,\rho_\perp,p_\perp',\epsilon_\perp')\,
\psi_\gamma(z,\rho_\perp,0_\perp,\epsilon_\perp)\label{wv}
\ee
\[
\hspace*{2cm}=\frac{N_ce^2e_q^2}{2(2\pi)^5}\int_{k_\perp,k_\perp'}
\mbox{tr}\Phi^\dagger(z,k_\perp',M,\epsilon_\perp')\Phi(z,k_\perp,M,
\epsilon_\perp)e^{i\rho_\perp(k_\perp'-k_\perp+zp_\perp')}\,,
\]
where we have used the notation of \cite{hks},
\be
\Phi(z,k_\perp,M,\epsilon_\perp)=\frac{1}{(k_\perp^2+M^2)}\left[
\,(1-z)\,\epsilon_\perp\cdot\sigma\,k_\perp\cdot\sigma-z\,k_\perp\cdot\sigma
\,\epsilon_\perp\cdot\sigma+iM\,\epsilon_\perp\cdot\sigma\,\right]\,,
\ee
$M$ is the quark mass and $\sigma_{1,2}$ are the first two Pauli matrices. 
Note that for $p_\perp'=0$, the average of the diagonal elements 
($\epsilon_\perp=\epsilon_\perp'$) in Eq.~(\ref{wv}) reproduces the well 
known formula for the square of the photon wave function \cite{nz}. 

Inserting Eq.~(\ref{wv}) into Eq.~(\ref{faf}) and introducing 
explicitly the required functionals $F_a$ specified by Eqs.~(\ref{dq}) and 
(\ref{dg}), the formulae of \cite{hks} for diffractive quark and gluon 
distribution are exactly reproduced.

\newpage

\newpage

\end{document}